\documentclass[10pt,journal,compsoc]{IEEEtran}

\usepackage[utf8]{inputenc}
\usepackage[T1]{fontenc}
\usepackage{graphicx}
\usepackage{tikz}
\usetikzlibrary{shapes.geometric, arrows.meta, positioning, fit, backgrounds, calc, shadows, decorations.pathreplacing}
\usepackage{pgfplots}
\pgfplotsset{compat=1.18}
\usepackage{booktabs}
\usepackage{listings}
\usepackage{xcolor}
\usepackage{hyperref}
\usepackage{cleveref}
\usepackage{balance}
\usepackage{tabularx}

\definecolor{policygreen}{HTML}{2E7D32}
\definecolor{evidenceblue}{HTML}{1565C0}
\definecolor{enforcered}{HTML}{C62828}
\definecolor{vaultyellow}{HTML}{F9A825}
\definecolor{neutralgray}{HTML}{616161}
\definecolor{lightbg}{HTML}{F5F5F5}
\definecolor{nistblue}{HTML}{003B6F}
\definecolor{eunavyblue}{HTML}{003399}
\definecolor{isopurple}{HTML}{6A1B9A}
\definecolor{oscalorange}{HTML}{E65100}
\definecolor{passgreen}{HTML}{388E3C}
\definecolor{failred}{HTML}{D32F2F}

\definecolor{yamlkey}{RGB}{0,80,160}
\definecolor{yamlvalue}{RGB}{160,50,0}
\definecolor{yamlcomment}{RGB}{110,110,110}
\definecolor{yamlbg}{RGB}{248,248,248}
\lstdefinelanguage{yaml}{
  keywords={name,value,control-id,description,props},
  keywordstyle=\color{yamlkey}\bfseries,
  sensitive=true,
  comment=[l]{\#},
  commentstyle=\color{yamlcomment}\itshape,
  stringstyle=\color{yamlvalue},
  morestring=[b]",
}
\lstdefinestyle{yaml}{
  language=yaml,
  basicstyle=\footnotesize\ttfamily,
  columns=fullflexible,
  showstringspaces=false,
  breaklines=true,
  frame=single,
  framerule=0.4pt,
  rulecolor=\color{gray!40},
  backgroundcolor=\color{yamlbg},
  xleftmargin=1.5em,
  framexleftmargin=1em,
  numberstyle=\tiny\color{gray},
  tabsize=2,
  literate={---}{{\textcolor{yamlkey}{-{}-{}-}}}3,
}

\begin{document}

\title{Making AI Compliance Evidence\\ Machine-Readable}

\author{
  Rodrigo~Cilla~Ugarte,
  Miguel~{\'A}ngel~Patricio~Guisado,
  Antonio~Berlanga~de~Jes{\'u}s,
  and~Jos{\'e}~Manuel~Molina~L{\'o}pez
  \IEEEcompsocitemizethanks{%
    \IEEEcompsocthanksitem R.~Cilla~Ugarte is with Ventural{\'\i}tica S.L., Donostia--San Sebasti{\'a}n, Spain.
      E-mail: rodrigo.cilla@venturalitica.ai
    \IEEEcompsocthanksitem M.~A.~Patricio, A.~Berlanga, and J.~M.~Molina are with the Applied Artificial Intelligence Group (GIAA), Computer Science Department, Universidad Carlos~III de Madrid, Legan{\'e}s, Spain.
    \IEEEcompsocthanksitem Preprint. This work has been submitted to \emph{IEEE Computer} (Special Issue on AI Governance and Compliance) for possible publication. Copyright may be transferred without notice, after which this version may no longer be accessible.
  }%
}

\IEEEtitleabstractindextext{
\begin{abstract}
AI Assurance---producing the machine-readable evidence required to demonstrate compliance with AI governance frameworks---has mature policy scaffolding but lacks the infrastructure to operationalize it. Organizations building high-risk AI systems under the EU AI Act face a gap: frameworks such as the EU AI Act, ISO/IEC~42001, and NIST AI RMF specify what to assure but provide no executable format for how. This paper proposes OSCAL---the NIST standard adopted for FedRAMP cybersecurity compliance---as a candidate interchange format for AI governance, complementing rather than replacing the emerging JTC21 standards stack. We define 16~property extensions covering lifecycle phases, enforcement semantics, risk traceability, and risk-acceptance justification, and present a three-layer Compliance-as-Code architecture (policy, evidence, enforcement) that generates assurance evidence as a byproduct of model training. The SDK produces native OSCAL Assessment Results validated against the NIST JSON schema. We test the approach on two Annex~III high-risk systems: a credit scoring model and a medical imaging segmentation system. The architecture and reference implementation are open-source under Apache~2.0.
\end{abstract}

}

\maketitle

\IEEEdisplaynontitleabstractindextext

\setcounter{secnumdepth}{0}

\begin{table*}[!t]
\centering
\caption{Cross-framework mapping of AI Assurance obligations. Each row shows how the same concern is expressed across three frameworks with different terminology and evidence expectations. No executable format unifies them.}
\label{tab:framework-mapping}
\small
\begin{tabularx}{\textwidth}{llXXX}
\toprule
\textbf{Concern} & \textbf{EU AI Act} & \textbf{Requirement} & \textbf{ISO/IEC 42001} & \textbf{NIST AI RMF} \\
\midrule
Risk management & Art.~9 & Continuous risk system across lifecycle & A.5.3 Risk treatment & Manage 2.2 \\
Data governance & Art.~10 & Quality criteria; bias examination & A.7.4 Data quality & Measure 2.6 \\
Technical docs & Art.~11 & Annex~IV documentation pre-market & A.6.2.4 Activity records & Govern 1.3 \\
Logging & Art.~12 & Automatic, traceable records & A.6.2.4 Activity records & Govern 1.3 \\
Transparency & Art.~13 & Information to deployers & A.8 Transparency & Map 3.5 \\
Human oversight & Art.~14 & Override capability; automation bias & A.8 Transparency & Govern 6.3 \\
Accuracy \& robustness & Art.~15 & Demonstrated performance; cybersecurity & A.5.4 Residual risk & Measure 2.7 \\
\bottomrule
\end{tabularx}
\end{table*}

\section{Introduction}
\label{sec:intro}

\textbf{\textit{AI governance frameworks tell organizations what to assure but not how to produce the evidence. We adapt OSCAL---the standard that made compliance-as-code possible in cybersecurity---to the AI domain, and validate the approach on two high-risk systems under the EU AI Act.}}

\medskip

AI Assurance---the machinery for producing the evidence that an AI system complies with a regulatory or management framework---sounds straightforward until you try to do it. Trustworthiness in the broader ethical sense enters as an input to the thresholds, not as an output of the machinery. The EU AI Act (Regulation (EU) 2024/1689), ISO/IEC~42001:2023, the NIST AI RMF (AI~100-1), Singapore's Model Framework, and emerging regulations in Brazil, South Korea, and Canada all describe what organizations should assure~\cite{kshetri_economics_ai_gov_2024}. None of them provides an executable format for encoding controls, a standard for evidence artifacts, or a reference architecture for wiring assurance into an ML pipeline. Marino and Lane~\cite{marino_computational_compliance_2026} put it bluntly: compliance with AI regulations can only realistically be achieved computationally. Sovrano et al.~\cite{sovrano2025aiactsoftware} reach a similar conclusion from the software-engineering side: drafting AI Act technical documentation is itself a software problem.

In practice, most teams treat compliance as a documentation exercise---risk registers, slide decks, manually authored reports---disconnected from the code that actually trains the model. The result is that when an auditor asks for proof, the evidence is six months old.

Cybersecurity solved a similar problem. When FedRAMP mandated OSCAL for cloud service authorizations, the enabling technology was OSCAL~\cite{nist_oscal_2025}---a NIST standard that turns security controls into version-controlled configuration and evidence into structured artifacts. RegTech adoption reduced compliance costs by 40--70\%~\cite{kshetri_regtech_2023}. AI governance has no equivalent infrastructure---yet.

In this article we extend OSCAL with 16~properties for AI lifecycle assurance, build a three-layer architecture (policy, evidence, enforcement) that generates compliance evidence during model training, and test it on two EU AI Act Annex~III high-risk systems: a credit scoring model and a medical imaging segmentation pipeline. The reference implementation is open-source (Apache~2.0).\footnote{\url{https://github.com/Venturalitica/venturalitica-sdk}}

\section{Background}
\label{sec:background}

Consider a European bank deploying a credit scoring system. It must satisfy the EU AI Act Articles~9--15 (risk management, data governance, documentation, logging, transparency, human oversight, accuracy)---but also DORA (digital operational resilience), NIS2 (network and information security), and GDPR (automated decision-making under Art.~22). If ISO/IEC~42001-certified, the bank maintains Annex~A controls on top. If serving the US market, the NIST AI RMF adds another layer. A single incident can trigger four reporting regimes with different authorities, forms, and deadlines~\cite{graux_interplay_2025}. \Cref{tab:framework-mapping} illustrates the overlap for one concern---bias---across three of these frameworks: three names for the same thing, three documentation streams, zero interoperability.

The EU AI Act obligations are binding from August~2026 under Regulation (EU) 2024/1689 as amended by the 2025 Digital Omnibus package, and they are evidence requirements---the Act asks providers to \emph{demonstrate} compliance, not merely to \emph{claim} it~\cite{bartsch_governance_highrisk_2025, wagner_navigating_eu_ai_act_2024}. Peckham~\cite{peckham_ai_harms_2024} proposes an AI harms framework, but without an executable mechanism. The first harmonised standard, prEN~18286, is in public enquiry but not yet cited in the Official Journal; until it is, providers cannot claim ``presumption of conformity'' (Art.~40). ISO/IEC~42001 certification is the strongest signal of due diligence available today, but it provides organizational governance---not the technical, machine-readable evidence that Arts.~9--15 demand.

Existing approaches address parts of this problem but not the whole. Model Cards~\cite{mitchell_model_cards_2019} and FactSheets~\cite{arnold_factsheets_2019} proposed structured documentation for AI transparency---static, single-framework artifacts, not continuous evidence streams tied to training runs. Governance platforms such as VerifyWise,\footnote{\url{https://verifywise.ai}} Modulos, and Credo~AI provide multi-tenant tooling, RBAC, attestation workflows, and regulator-facing dashboards---features outside the scope of an SDK---but each produces evidence in platform-specific shapes that are not directly portable across jurisdictions. The missing layer is the interchange format: continuous, machine-readable evidence generated as a byproduct of the ML pipeline, in a standardized shape that any auditor or regulatory tool can verify. That is what we build.

\section{OSCAL as Lingua Franca for AI Governance}
\label{sec:oscal}

If compliance-as-code is the goal, what language should the code be written in? For cybersecurity, the answer is OSCAL~\cite{nist_oscal_2025}---the Open Security Controls Assessment Language, developed by NIST. OSCAL provides six models covering the full compliance lifecycle:

\begin{itemize}
  \item \textbf{Catalog}: a library of controls (e.g., ``bias must not exceed threshold $X$'').
  \item \textbf{Profile}: a selection of controls from one or more catalogs applicable to a specific context (e.g., ``controls for a credit scoring system under EU AI Act Annex~III'').
  \item \textbf{Component definition}: a description of how a system or tool implements specific controls.
  \item \textbf{Assessment plan}: what will be tested, when, and against which criteria.
  \item \textbf{Assessment results}: the outcome of evaluating controls---pass, fail, values, timestamps.
  \item \textbf{Plan of Action and Milestones} (POA\&M): remediation tracking for failed controls.
\end{itemize}

All six models are serializable as JSON, XML, or YAML, enabling version control alongside source code. FedRAMP mandates OSCAL for all cloud service authorizations.

For AI governance, the equivalent does not exist. There are no OSCAL catalogs for the EU AI Act, no profiles for high-risk systems, no assessment plans encoding Articles~9--15. We built the first one.

\subsection{An OSCAL Profile for AI Assurance}

OSCAL's \texttt{implemented-requirement} structure holds a control ID, a description, and a list of typed properties. The standard needs 16~new properties to work for AI (\Cref{tab:oscal-extensions}). Together they express \emph{what} to evaluate, \emph{when} and \emph{how}, \emph{why}---linking each technical check to its originating risk, treatment plan, and organizational objective---and \emph{on what basis} the thresholds and acceptance criteria are defensible.

\begin{table*}[t]
\centering
\caption{Proposed OSCAL property extensions for AI Assurance, with normative justification across three frameworks}
\label{tab:oscal-extensions}
\small
\begin{tabularx}{\textwidth}{lXlll}
\toprule
\textbf{Property} & \textbf{Purpose} & \textbf{EU AI Act} & \textbf{ISO 42001} & \textbf{NIST RMF} \\
\midrule
\texttt{metric\_key} & Metric function identifier & 10.2(f), 15.1 & A.6.2.3, A.7.4 & Measure 2.6 \\
\texttt{operator} & Comparison operator & 9.2(b), 15.1 & A.6.2.3 & Measure \\
\texttt{threshold} & Quantitative threshold & 9.2(b), 15.1 & A.5.3 & Measure \\
\texttt{severity} & Control criticality & 9.8 & A.5.3 & Map 2.2 \\
\texttt{lifecycle\_phase} & One or more of training, validation, monitoring, incident & 9--15, 72, 73 & A.6.2 & Govern 1.7 \\
\texttt{enforcement\_mode} & Monitor, warn, or block & 14.4(e), 72, 73 & A.6.2.6, A.8.3 & Manage 2 \\
\texttt{evaluation\_method} & Automated, manual, hybrid & 9.9, 43 & A.6.2.3 & Measure 4 \\
\texttt{evaluation\_window} & Per-run, periodic, sliding & 9.3, 72 & A.6.2.6 & Manage 3 \\
\texttt{target\_type} & System, dataset, or model & 10, 11, 15 & A.4.2, A.7 & Map 1 \\
\texttt{risk\_id} & Link to identified risk & 9.2(a) & \S6.1.2 & Map 3 \\
\texttt{treatment\_id} & Link to risk treatment & 9.4 & \S6.1.3 & Manage 2 \\
\texttt{policy\_id} & Link to governing policy & 9 & \S5.2, A.2 & Govern 1.1 \\
\texttt{objective\_id} & Link to AI objective & --- & \S6.2 & Govern 1.3 \\
\texttt{risk\_acceptance\_criteria} & Explicit acceptance criterion for residual risk & 9.5 & \S6.1.2 & Manage 1.3 \\
\texttt{threshold\_justification} & Citation/rationale for the numeric threshold & 9.2(b), 10 & A.5.3, A.7.4 & Measure 2.9 \\
\texttt{stakeholder\_consultation\_ref} & Record of stakeholder deliberation behind the threshold & 10.2(d) & \S4.2, A.3.2 & Map 1.6 \\
\bottomrule
\end{tabularx}
\end{table*}

The 16~properties form a single, coherent profile specification---the primary contribution of this section. Of these, the last three respond to a question Art.~9.5 forces explicitly: \emph{why is this threshold defensible?} Making \texttt{threshold\_justification} and \texttt{risk\_acceptance\_criteria} first-class properties obliges the evidence chain to name its normative source (for example, the EEOC Four-Fifths Rule, a WHO guideline, or a documented stakeholder deliberation referenced via \texttt{stakeholder\_consultation\_ref}).

The profile is designed so that different consumers use the relevant subset: the training-time SDK enforces controls tagged \texttt{lifecycle\_phase: training} or \texttt{validation} using \texttt{metric\_key}, \texttt{operator}, \texttt{threshold}, \texttt{severity}, and \texttt{enforcement\_mode}; a runtime proxy (discussed in \S Discussion) consumes controls tagged \texttt{monitoring} together with \texttt{evaluation\_window}; and a governance platform reads the traceability chain to close the loop with organizational risk management. All coexist in the same OSCAL file; unknown properties are ignored, preserving backward compatibility.

The specification is fully documented here for adoption by any implementation. The open-source reference implementation produces OSCAL Assessment Results documents validated against the official NIST JSON schema (v1.2.1), demonstrating that the profile is not only specified but executable.

\subsection{Why Not a New Format?}

We extend OSCAL rather than designing a new format for three reasons. First, OSCAL already has validation libraries, converters, and institutional adoption through FedRAMP. Second, NIST maintains both the AI RMF and OSCAL~\cite{nist_oscal_2025}; a future OSCAL Profile for AI is a plausible evolutionary step. Third, a single OSCAL file can reference controls from the EU AI Act, NIST AI RMF, and ISO/IEC~42001 simultaneously, eliminating the triple-documentation problem. OSCAL supplies an interoperability layer---a machine-readable envelope---for the requirements that other governance instruments articulate in prose.

\subsection{Lifecycle Semantics}

The EU AI Act distinguishes between pre-market obligations (Arts.~9--15, before a system is placed on the market) and post-market obligations (Art.~72, continuous monitoring; Art.~73, incident reporting). A single assurance control---for example, ``demographic parity difference $<$ 0.10''---applies across both regimes, but with different semantics at each lifecycle phase.

In OSCAL terms, each phase is expressed by repeating the \texttt{lifecycle\_phase} property on a single \texttt{implemented-requirement}: a control that runs both at training and at runtime carries \texttt{lifecycle\_phase: training} and \texttt{lifecycle\_phase: monitoring}, and each consumer picks up the subset relevant to its scope. The phase determines \emph{what the check operates on}: \texttt{training} controls evaluate raw training data (class imbalance, demographic representation --- Art.~10); \texttt{validation} controls evaluate model predictions (accuracy, fairness of decisions --- Art.~15); \texttt{monitoring} controls evaluate production inference traffic (drift, fairness over time --- Art.~72); \texttt{incident} controls feed into the Art.~73 reporting path.

The \texttt{evaluation\_window} property applies primarily to the monitoring phase, where controls run continuously rather than per-pipeline: \texttt{per-run} is the default at training and validation, whereas \texttt{periodic} and \texttt{sliding-window} are consumed by the runtime proxy. The \texttt{enforcement\_mode} determines the consequence: \texttt{block} halts the pipeline by raising an exception on the failing control, \texttt{warn} emits an alert without interrupting execution, and \texttt{monitor} logs silently for audit. A single OSCAL catalog---authored once by a compliance team---therefore drives both the training-time SDK and the runtime proxy, with each consumer selecting the controls whose \texttt{lifecycle\_phase} matches its scope.

\subsection{Traceability Chain}

The traceability properties enable a vertical chain that mirrors the full ISO/IEC~42001 PDCA hierarchy. The chain flows downward from organizational governance to technical evidence: the AI Policy (\S5.2) sets the direction; AI Objectives (\S6.2) operationalize it; Risk Assessment (\S6.1.2) identifies threats; Risk Treatment (\S6.1.3) prescribes mitigations; Assurance Controls (Annex~A) implement them as measurable checks. The chain continues upward in the \emph{Check} and \emph{Act} phases: each OSCAL Assessment Result populates performance evaluation (\S9.1), and each POA\&M item populates corrective action (\S10.1).

Each OSCAL control carries typed links to every level above it: \texttt{policy\_id}, \texttt{objective\_id}, \texttt{risk\_id}, \texttt{treatment\_id}. When a metric fails---a disparate impact ratio of 0.65 against a 0.80 threshold---the chain traces upward to the treatment (``apply group-aware reweighting''), to the risk (``gender discrimination in credit approval''), to the objective (``ensure non-discriminatory credit decisions''), and to the policy (``Four-Fifths Rule per EEOC guidelines''). This is what Art.~9.2(a) requires.

\section{Architecture}
\label{sec:architecture}

\begin{figure*}[t]
\centering
\begin{tikzpicture}[
  layer/.style={rounded corners=6pt, minimum width=15.5cm, thick, inner sep=8pt},
  ltitle/.style={font=\sffamily\bfseries\footnotesize, anchor=north west},
  lsub/.style={font=\sffamily\tiny, text=neutralgray, anchor=north west},
  pfile/.style={draw=#1!60, fill=#1!4, rounded corners=3pt, font=\sffamily\scriptsize, minimum height=0.7cm, thick, inner xsep=6pt},
  probe/.style={draw=evidenceblue!50, fill=white, rounded corners=3pt, minimum width=1.5cm, minimum height=0.85cm, font=\sffamily\tiny, align=center},
  artlabel/.style={font=\sffamily\tiny\bfseries, text=evidenceblue!70},
  outcome/.style={rounded corners=3pt, font=\sffamily\scriptsize\bfseries, minimum width=1.8cm, minimum height=0.5cm},
  flow/.style={-{Stealth[length=2.5mm]}, thick, neutralgray!60},
  flowlabel/.style={font=\sffamily\tiny, text=neutralgray},
]

\node[layer, draw=policygreen!60, fill=policygreen!3, minimum height=1.8cm] (L1) at (0,5.5) {};
\node[ltitle, text=policygreen!80!black] at (-7.4,6.2) {POLICY};
\node[lsub] at (-7.4,5.85) {OSCAL assessment plans, version-controlled in git};
\node[pfile=policygreen] at (-3.5,5.2) {\texttt{data\_policy.oscal.yaml} {\tiny(Art.~10)}};
\node[pfile=policygreen] at (3.5,5.2) {\texttt{model\_policy.oscal.yaml} {\tiny(Art.~15)}};

\node[layer, draw=evidenceblue!50, fill=evidenceblue!2, minimum height=2.2cm] (L2) at (0,2.8) {};
\node[ltitle, text=evidenceblue!80!black] at (-7.4,3.7) {EVIDENCE};
\node[lsub] at (-7.4,3.35) {7 probes activated by \texttt{with vl.monitor():}};

\node[probe] (p1) at (-5.8,2.4) {\textbf{Trace}\\code AST};
\node[probe] (p2) at (-3.8,2.4) {\textbf{Artifact}\\SHA-256};
\node[probe] (p3) at (-1.8,2.4) {\textbf{BOM}\\CycloneDX};
\node[probe] (p4) at (0.2,2.4) {\textbf{Integrity}\\env hash};
\node[probe] (p5) at (2.2,2.4) {\textbf{Hardware}\\CPU/GPU};
\node[probe] (p6) at (4.2,2.4) {\textbf{Carbon}\\kgCO\textsubscript{2}};
\node[probe] (p7) at (6.2,2.4) {\textbf{Handshake}\\enforce?};

\node[artlabel] at (-4.8,1.75) {Arts.~10--12};
\node[artlabel] at (-0.8,1.75) {Arts.~11, 15};
\node[artlabel] at (3.2,1.75) {Art.~11};
\node[artlabel] at (6.2,1.75) {Art.~9};

\node[layer, draw=enforcered!50, fill=enforcered!2, minimum height=1.8cm] (L3) at (0,0.2) {};
\node[ltitle, text=enforcered!80!black] at (-7.4,0.9) {ENFORCEMENT};
\node[lsub] at (-7.4,0.55) {\texttt{vl.enforce()}: OSCAL controls $\rightarrow$ metric function registry};

\node[outcome, draw=passgreen!60, fill=passgreen!8, text=passgreen!80!black] at (-3,-0.3) {PASS $\rightarrow$ Finding: satisfied};
\node[outcome, draw=failred!60, fill=failred!8, text=failred!80!black, minimum width=4cm] at (2.5,-0.3) {FAIL $\rightarrow$ Risk + POA\&M item};

\node[draw=vaultyellow!60, fill=vaultyellow!5, rounded corners=4pt, minimum width=8cm, minimum height=0.7cm, thick, font=\sffamily\scriptsize] (vault) at (0,-1.7) {\textbf{OSCAL Assessment Results} ~~\textbar~~ \texttt{assessment-results.oscal.json}};

\draw[flow] (L1.south) -- node[right, flowlabel] {controls loaded} (L2.north);
\draw[flow] (L2.south) -- node[right, flowlabel] {observations} (L3.north);
\draw[flow, dashed, vaultyellow!70!black] (L3.south) -- node[right, flowlabel] {findings + risks} (vault.north);

\end{tikzpicture}
\caption{Three-layer architecture. OSCAL policies define controls with the proposed AI extensions (including metric selection and thresholds). Seven probes collect evidence during training. The enforcement engine dispatches each control to a metric function from the SDK's extensible registry (fairness, privacy, data quality, performance) and produces OSCAL Assessment Results---with auto-generated risks and POA\&M items for failed controls.}
\label{fig:architecture}
\end{figure*}
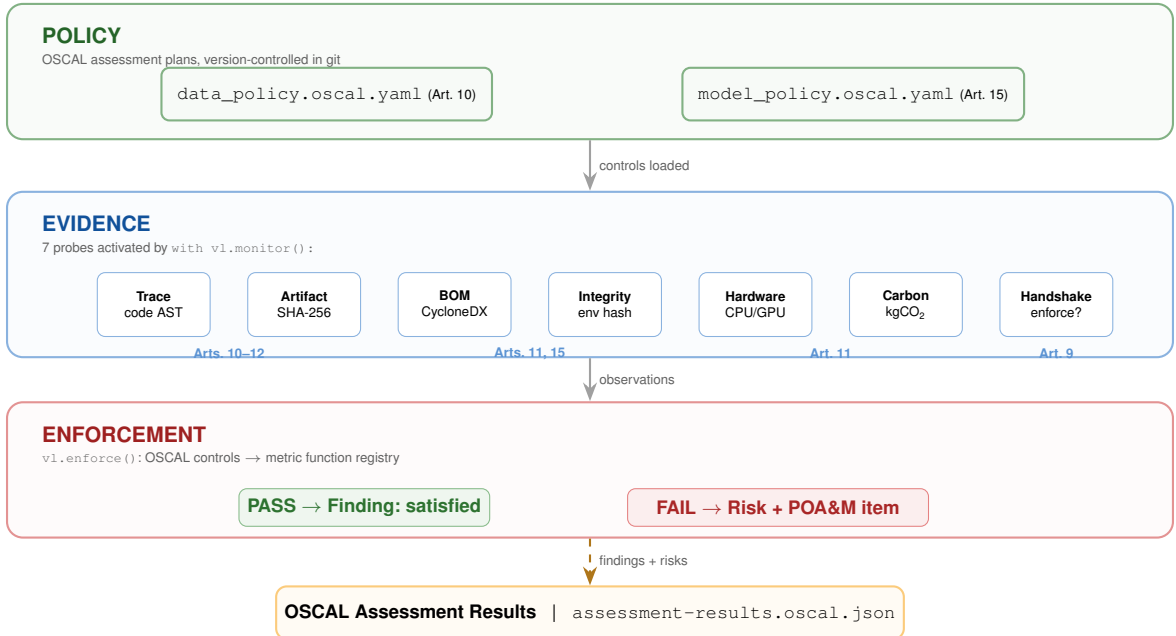

\Cref{fig:architecture} shows how the pieces fit together. The architecture has three layers, all focused on \emph{pre-market placement}---generating the evidence that Articles~9--15 require before the system reaches the market. Post-market monitoring (Art.~72) and incident reporting (Art.~73) run on a separate infrastructure layer and are outside the scope of this work.

\subsection{Layer 1: Policy (OSCAL)}

Compliance controls are expressed as OSCAL assessment plan files, version-controlled alongside model source code. We adopt a separation between \emph{data policies} (evaluated before training, addressing Art.~10) and \emph{model policies} (evaluated after training, addressing Art.~15). Each control \emph{references} a metric function (from the SDK's library of 104~functions), a comparison operator, a quantitative threshold, and the proposed AI extensions (lifecycle phase, enforcement mode, severity, risk traceability). OSCAL defines the control intent and acceptance criteria; the metric library provides the computational implementation.

\Cref{fig:listing} shows an example control that checks gender disparate impact using the Four-Fifths Rule, applied during the training phase, with blocking enforcement linked to an identified risk. The choice of a single threshold over a specific metric is deliberate; Bell et al.~\cite{bell2023fairness} show that in practice fairness criteria previously argued to be incompatible can be satisfied with bounded relaxations, and the OSCAL profile lets those relaxations be declared explicitly per control rather than assumed.

\begin{figure}[t]
\centering
\begin{lstlisting}[style=yaml, label={lst:oscal-example}]
assessment-plan:
  metadata:
    title: "Art. 10 Data Governance"
  control-implementations:
    - implemented-requirements:
      - control-id: credit-data-bias
        description: "Gender disparate impact
          follows Four-Fifths Rule"
        props:
          - name: metric_key
            value: disparate_impact
          - name: operator
            value: ">="
          - name: threshold
            value: "0.8"
          - name: threshold_justification
            value: "EEOC Four-Fifths Rule (1978)"
          - name: risk_acceptance_criteria
            value: "residual DI >= 0.80
              after mitigation"
          - name: severity
            value: high
          - name: lifecycle_phase
            value: training
          - name: enforcement_mode
            value: block
          - name: risk_id
            value: R-042
          - name: treatment_id
            value: T-017
\end{lstlisting}
\vspace{-0.5em}
\caption{Excerpt from a real OSCAL policy file (abridged for space). The first three properties drive metric evaluation; \texttt{threshold\_justification} and \texttt{risk\_acceptance\_criteria} name the normative source and residual-risk bound; \texttt{severity} labels criticality for reporting; \texttt{lifecycle\_phase} and \texttt{enforcement\_mode} control which engine evaluates the check and how it reacts to failure; \texttt{risk\_id} and \texttt{treatment\_id} link to the risk register. Full examples at \url{https://github.com/Venturalitica/venturalitica-sdk-samples}.}
\label{fig:listing}
\end{figure}

\subsection{Layer 2: Evidence Collection}

The evidence layer wraps the ML pipeline in a context manager that activates seven concurrent probes---covering code analysis (AST trace), data integrity (SHA-256 hashes), software supply chain (CycloneDX BOM~\cite{nocera_aibom_mlr_2025}), environment fingerprinting, hardware telemetry, carbon emissions, and enforcement verification---each mapped to a specific EU AI Act article (Arts.~9--12, 15). Research documents systematic gaps in ML evidence capture; the probes address them.

Concretely, a developer wraps their training code in \texttt{with vl.monitor("my-run"):} and, inside the block, calls \texttt{vl.enforce(data=df, policy="my-policy.oscal.yaml")}. The \texttt{monitor} context activates all probes in parallel; \texttt{enforce} loads the YAML policy, evaluates each control against the data, and caches results for the OSCAL artifact generated at context exit. The AST probe traces which library functions the training script invokes; the Integrity probe computes SHA-256 hashes of input data splits and output model weights; the BOM probe introspects the Python environment and generates a CycloneDX bill of materials; and the Environment probe records CUDA version, GPU model, and OS fingerprint. No modification of the training code itself is required.

Evidence is persisted to a local vault (\texttt{.venturalitica/runs/\{run\_id\}/}) as a native OSCAL Assessment Results document (\texttt{assessment-results.oscal.json}), validated against the NIST JSON schema (v1.2.1). When controls fail, the SDK generates an OSCAL Plan of Action and Milestones (\texttt{poam.oscal.json}) with open risk items linked to the failing findings.

\subsection{Layer 3: Enforcement}

The enforcement engine reads the OSCAL controls and dispatches each to the corresponding metric function from the SDK's extensible registry covering fairness, privacy, data quality, and performance. The OSCAL control defines \emph{what} to measure and the acceptance threshold; the metric function defines \emph{how} to compute it. For each control, the engine resolves \texttt{metric\_key} to a function, loads the relevant data or predictions, computes the metric, and compares the result against \texttt{operator} + \texttt{threshold}. Each evaluation produces an OSCAL Observation; each control outcome becomes a Finding (\texttt{satisfied} or \texttt{not-satisfied}). Failures auto-generate an OSCAL Risk with characterization facets---the metric name, actual value, threshold, operator, and affected demographic groups---plus a POA\&M remediation item. The \texttt{enforcement\_mode} property governs the response: \texttt{monitor} logs only, \texttt{warn} emits an alert, \texttt{block} halts the pipeline by raising an exception at the failing control.

\section{Validation}
\label{sec:validation}

We tested the architecture on two AI systems classified as high-risk under EU AI Act Annex~III.

\subsection{Scenario A: Credit Scoring (Annex III, Area 5b)}

Credit scoring systems are explicitly listed in Annex~III, Area~5(b). We use the UCI German Credit dataset~\cite{uci_german_credit} (1,000~samples, 20~features, binary target) to demonstrate the full assurance lifecycle. The scenario implements a logistic regression classifier with group-aware sample reweighting as a bias mitigation strategy, wrapped in the SDK's \texttt{monitor()} context manager. The complete scenario---including OSCAL policies, training script, and generated evidence---is available as an open-source reproducibility package.\footnote{\url{https://github.com/Venturalitica/venturalitica-scenario-financial}}

The end-to-end flow maps directly onto the three-layer architecture. A compliance officer authors an OSCAL policy (\texttt{credit-scoring.oscal.yaml}) with five controls: three pre-training (class imbalance ratio, gender disparate impact, age disparate impact) and two post-training (accuracy, gender demographic parity difference). The developer wraps the training script in \texttt{with vl.monitor("credit-scoring"):} and calls \texttt{vl.enforce(data=df, policy="credit-scoring.oscal.yaml")} inside. The evidence layer activates all probes; the enforcement layer evaluates the five controls against the data and predictions.

\Cref{tab:results} summarizes the results. The age disparity fails (0.286 vs.\ 0.50 threshold); with \texttt{enforcement\_mode: block}, training would halt. After group-aware sample reweighting ($w_{g,y} = P(g) \cdot P(y) / P(g, y)$), gender demographic parity drops to 0.012 while accuracy holds at 0.795.

\begin{table}[h]
\centering
\small
\begin{tabular}{llcccl}
\toprule
\textbf{Phase} & \textbf{Control} & \textbf{Actual} & \textbf{Op} & \textbf{Thresh.} & \textbf{Result} \\
\midrule
Pre & Class imbalance & 0.429 & $>$ & 0.20 & \textsc{pass} \\
Pre & Gender DI & 0.818 & $>$ & 0.80 & \textsc{pass} \\
Pre & Age DI & 0.286 & $>$ & 0.50 & \textsc{fail} \\
\midrule
Post & Accuracy & 0.795 & $\geq$ & 0.70 & \textsc{pass} \\
Post & Gender DP diff & 0.012 & $<$ & 0.10 & \textsc{pass} \\
\bottomrule
\end{tabular}
\caption{Pre-training (Art.~10) and post-training (Art.~15) audit results for credit scoring. DI = Disparate Impact, DP = Demographic Parity.}
\label{tab:results}
\end{table}

The age disparity failure triggers an OSCAL Risk in the assessment results, with characterization facets recording the metric (\texttt{disparate\_impact}), actual value (0.286), threshold (0.50), and operator (\texttt{gt}). The POA\&M auto-generates an open remediation item linked to this risk. The enforcement engine also reports per-group breakdowns: female positive rate 35.16\%, male positive rate 27.68\%, ratio 0.787---granularity that an auditor needs to verify the underlying distribution, not just a pass/fail flag.

After group-aware sample reweighting ($w_{g,y} = P(g) \cdot P(y) / P(g, y)$), the post-training audit shows gender demographic parity dropped to 0.012 while accuracy held at 0.795. The age disparity persists because the mitigation targeted gender only; a subsequent cycle would address it before market placement.

The evidence vault for this run contains: the OSCAL Assessment Results document (3~observations, 3~findings, 1~risk for pre-training; 2~findings, 0~risks for post-training), the POA\&M with one open item, a trace file with AST analysis of the training script, SHA-256 hashes of data splits, a CycloneDX BOM (scikit-learn~1.3.0, 12~dependencies), hardware metadata (4~CPUs, 2.1~GB peak RAM), and carbon emissions (0.002~kgCO$_2$). All artifacts are OSCAL-compliant and validated against the NIST JSON schema.

\subsection{Scenario B: Medical Imaging (Annex III, Area 5a)}

AI systems used as safety components in medical devices are high-risk under Annex~III, Area~5(a), and additionally subject to the Medical Device Regulation (EU)~2017/745. We apply the same architecture to a 3D whole-body CT segmentation pipeline built on MONAI~1.3 with a SegResNet backbone pre-trained to segment 105~anatomical structures. Unlike the credit-scoring scenario, here the model is downloaded as published weights---representative of the manufacturer-as-deployer case: a hospital or SaMD vendor that integrates a third-party model and must still demonstrate EU AI Act conformity and ISO~42001 alignment on its own deployment. The \texttt{monitor()} context manager activates all seven probes without modifying the inference code.\footnote{\url{https://github.com/Venturalitica/venturalitica-scenario-medical}}

The OSCAL policy (\texttt{medical-ct.oscal.yaml}) defines controls for Dice coefficient, sensitivity, and specificity---each stratified by patient cohort (age, gender). The developer wraps the MONAI inference loop in \texttt{with vl.monitor("medical-ct"):} and calls \texttt{vl.enforce(data=preds, policy="medical-ct.oscal.yaml")} inside; the probes capture model weights hashes, a CycloneDX BOM (MONAI + 47~dependencies), and CUDA environment metadata. The enforcement engine evaluates the controls across all cohort stratifications, producing 37~evaluation rows---each an OSCAL Observation traceable to its originating control. The BOM lists MONAI~1.3, PyTorch~2.1, and 47~transitive dependencies; the Integrity probe records the full computational environment (CUDA~12.1, A100~80GB), addressing Art.~11/Annex~IV documentation requirements.

The two scenarios cover the two canonical deployment modes: training your own model on tabular data (Scenario A) and deploying a pre-trained volumetric model (Scenario B). The pipeline code differs---scikit-learn vs MONAI/PyTorch---but the SDK integration is identical: the same \texttt{monitor()} call, the same OSCAL policy format, the same probes and enforcement engine. Only the policy content (metrics, thresholds, framework references) and whether \texttt{lifecycle\_phase} targets \texttt{training} or \texttt{validation} change between domains.

\subsection{Summary of Generated Evidence}

A single pipeline execution---whether training or inference---generates a compliance bundle: the OSCAL Assessment Results and POA\&M documents plus the probe artifacts (BOM, hashes, environment fingerprint, carbon trace), colocated in the run directory. The bundle is machine-readable, locally stored, and independently verifiable by an auditor.

\section{Discussion}
\label{sec:discussion}

\noindent\textbf{Implications for engineering teams.} The architecture shifts compliance from a recurring audit cost to an amortized pipeline cost: evidence is produced as a byproduct of the pipeline rather than reconstructed ex-post. Teams get immediate feedback on data quality and fairness (\Cref{tab:results}), enabling remediation before market placement rather than after a failed audit.

\textbf{Implications for regulators and auditors.} Machine-readable OSCAL evidence prepares the infrastructure for scalable conformity assessment. Regulators and notified bodies \emph{could} consume assessment results programmatically---verifying that controls were evaluated against thresholds at recorded timestamps---once the format is recognised in sector-specific guidance or a notified-body procedure; this recognition is a policy step beyond the technical contribution of this work. The same structured evidence feeds auditor dashboards: a viewer can render OSCAL Findings, Risks, and POA\&M items as an interactive risk register, surfacing failing controls and affected demographic groups without requiring the auditor to parse raw JSON. Art.~14 human oversight is a cognitive problem as much as a technical one~\cite{laux_automation_bias_2025}; standardized evidence formats make it tractable to build the review interfaces that overseers actually need.

\textbf{Implications for standards bodies.} The proposed extensions (\Cref{tab:oscal-extensions}) are a starting point for standardization. The CEN/CENELEC work on prEN~18286 could adopt an executable policy format alongside natural-language controls. A NIST OSCAL Profile for AI---analogous to existing FedRAMP profiles---could unify AI RMF sub-categories with EU AI Act articles in one catalog.

\textbf{From pre-market to post-market monitoring.} The SDK described here targets the pre-market obligations of Arts.~9--15, but the same OSCAL foundation extends naturally to Art.~72 (post-market monitoring) and Art.~73 (serious incident reporting). The \texttt{lifecycle\_phase} and \texttt{evaluation\_window} properties we introduced (\Cref{tab:oscal-extensions}) are the hooks a runtime layer uses. A hypothetical production-monitoring agent---deployable under Privacy-by-Design / Bring-Your-Own-Cloud constraints---would act as a second consumer of the same OSCAL catalog: ingesting the controls tagged \texttt{lifecycle\_phase:~monitoring} and emitting the same Assessment Results structure (observations, findings, risks, POA\&M). Continuous-evaluation semantics governed by \texttt{evaluation\_window} (including sliding-window assessment of drift and demographic parity over production traffic) follow naturally once that layer exists. Training-time and runtime enforcement would therefore share a single source of truth, not a duplicate governance stack.

\textbf{From classical ML to agentic AI.} Recent work proposes governance for autonomous AI agents~\cite{kshetri_governing_agentic_2025}, including trajectory-level compliance policies. Nannini et al.~\cite{nannini_ai_agents_eu_2026} argue that high-risk agentic systems with untraceable behavioral drift---quantified empirically by Rath~\cite{rath2026agentdrift}---cannot currently meet the AI Act's essential requirements; their three oversight modalities (retrospective, real-time, continuous) map to our \texttt{enforcement\_mode} values, and \texttt{lifecycle\_phase: monitoring} with \texttt{evaluation\_window: sliding} are the hooks a continuous layer needs. Runtime-governance frameworks such as MI9~\cite{wang2025mi9} and policies-on-paths~\cite{furman2026policiesonpaths} describe complementary mechanisms (telemetry, conformance state machines, trajectory policies) that a compliant runtime proxy would implement on top of the same OSCAL catalog. These proposals assume an assurance layer exists for the models agents invoke. Our architecture is that prerequisite: assure the models first, then assure the agents.

\textbf{Limitations.} The architecture covers Arts.~9--15 but not organizational requirements (Arts.~16--17). The multi-regulation problem~\cite{graux_interplay_2025} is beyond scope. Validation is limited to tabular and volumetric imaging scenarios; NLP, LLM, and recommender systems remain future work, as do GPAI models with systemic risk (Arts.~51--55), which fall under a distinct regime outside this profile. OSCAL JSON is designed for engineers and regulators---a parallel human-readable layer is needed for public accountability.

\section{Conclusion}
\label{sec:conclusion}

AI governance frameworks tell you what to measure. OSCAL, extended with the 16~properties we describe here, tells the machine how to measure it.

We tested the approach on a credit scoring model and a medical imaging pipeline---two different modalities, same OSCAL policies, same probes, same enforcement engine. The SDK produces native Assessment Results validated against the NIST JSON schema. When a control fails, it generates a Plan of Action and Milestones with the risk linked to the specific finding. The evidence is machine-readable, version-controlled, and auditor-ready---produced as a byproduct of training, not as a separate exercise.

The takeaway for engineering teams: compliance is an infrastructure problem. For regulators: machine-readable evidence scales. For standards bodies working on prEN~18286: the OSCAL extensions proposed here are a concrete starting point.

\subsection*{Declaration of Competing Interests}
R.~Cilla~Ugarte is founder of Ventural{\'\i}tica S.L., the company that develops and maintains the open-source \texttt{venturalitica-sdk} described in this work. The remaining authors declare no competing interests. This manuscript received no external funding from Ventural{\'\i}tica S.L. or any other organization; the reference implementation is released under the Apache~2.0 license and is independently verifiable.

\subsection*{AI Use Declaration}
Claude (Anthropic, Opus~4 model via Claude Code CLI) was used during manuscript preparation for: (1)~literature search assistance in the Background section (identifying and cross-referencing related work); (2)~LaTeX formatting and table layout across all sections; and (3)~prose editing for grammar and clarity. All scientific content---the OSCAL extension design, the three-layer architecture, experimental design, and analysis of results---was authored solely by the authors. All AI-assisted outputs were reviewed, verified, and revised by the authors, who retain full responsibility for the accuracy and integrity of the work.

\balance

\bibliographystyle{IEEEtran}
\bibliography{references}


\begin{IEEEbiography}[]{Rodrigo Cilla Ugarte}
is founder of Venturalítica S.L. and holds a PhD in Computer Science from Universidad Carlos III de Madrid. He previously led CE marking for medical device software at ULMA Medical Technologies and is a certified ISO/IEC~42001 Lead Auditor. Contact: rodrigo.cilla@venturalitica.ai.
\end{IEEEbiography}

\begin{IEEEbiography}[]{Miguel Ángel Patricio Guisado}
is a Full Professor at Universidad Carlos III de Madrid (GIAA group), with a PhD from Universidad Politécnica de Madrid. His research covers computer vision, machine learning, and distributed systems.
\end{IEEEbiography}

\begin{IEEEbiography}[]{Antonio Berlanga de Jesús}
is an Associate Professor at Universidad Carlos III de Madrid (GIAA group), with a PhD in Computer Engineering from UC3M. His research covers machine learning, multi-objective optimization, and decision-making.
\end{IEEEbiography}

\begin{IEEEbiography}[]{José Manuel Molina López}
is a Full Professor at Universidad Carlos III de Madrid and Director of the GIAA group, with a PhD in Telecommunications Engineering from UPM. His research spans machine learning, deep learning, and multi-agent systems.
\end{IEEEbiography}

\end{document}